\documentclass[%
 aip,
 amsmath,amssymb,
 reprint,%
]{revtex4-1}

\usepackage{graphicx}
\usepackage{dcolumn}
\usepackage{bm}

\usepackage[T1]{fontenc}

\begin{document}

\preprint{J. Appl. Phys.}

\title[Superconductivity in V$_{1-x}$Zr$_x$ alloys]{Evolution of high field superconductivity and high critical current density in the as-cast V$_{1-x}$Zr$_x$ alloys}

\author{L. S. Sharath Chandra}
\email{lsschandra@rrcat.gov.in}
 \affiliation{Free Electron Laser Utilization Laboratory, Raja Ramanna Centre for Advanced Technology, Indore - 452 013, India.}
 \author{Sabyasachi Paul}%
\affiliation{Free Electron Laser Utilization Laboratory, Raja Ramanna Centre for Advanced Technology, Indore - 452 013, India.}%
\affiliation{Homi Bhabha National Institute, Training School Complex, Anushakti Nagar, Mumbai 400 094, India.}%
\author{Ashish Khandelwal}
\affiliation{Free Electron Laser Utilization Laboratory, Raja Ramanna Centre for Advanced Technology, Indore - 452 013, India.}%
\author{Archna Sagdeo}
\affiliation{Hard X-ray Applications Laboratory, Raja Ramanna Centre for Advanced Technology, Indore - 452 013, India.}%
\affiliation{Homi Bhabha National Institute, Training School Complex, Anushakti Nagar, Mumbai 400 094, India.}%
\author{R. Venkatesh}
\affiliation{UGC-DAE Consortium for Scientic Research, Takshashila Campus, Khandwa Road, Indore 452 017, India.}%
\author{Kranti Kumar}
\affiliation{UGC-DAE Consortium for Scientic Research, Takshashila Campus, Khandwa Road, Indore 452 017, India.}%
\author{A. Banerjee}
\affiliation{UGC-DAE Consortium for Scientic Research, Takshashila Campus, Khandwa Road, Indore 452 017, India.}%
\author{M. K. Chattopadhyay}
\affiliation{Free Electron Laser Utilization Laboratory, Raja Ramanna Centre for Advanced Technology, Indore - 452 013, India.}%
\affiliation{Homi Bhabha National Institute, Training School Complex, Anushakti Nagar, Mumbai 400 094, India.}%

\date{\today}

\begin{abstract}
We report here the structural, electrical and magnetic properties of as-cast V$_{1-x}$Zr$_x$ alloys ($x$ =0 - 0.4) at low temperatures. We observe that all the alloys undergo successive peritectic and eutectic reactions during cooling from the melt which leads to the formation of five phases, namely, a body centred cubic $\beta$-V phase, two phases with slightly different compositions having face centred cubic ZrV$_2$ structure, a hexagonal closed packed $\alpha$-Zr phase, and the $\beta$-Zr precipitates. The amount of each phase is found to be dependent on the concentration of zirconium in vanadium. The $\beta$-V and ZrV$_2$ phases show superconductivity below 5.3~K and 8.5~K respectively. We show that the critical current density is large for V-rich V$_{1-x}$Zr$_x$ alloys with $x >$ 0.1. The grain boundaries generated from the eutectic reaction, and the point defects formed due to the variation in the composition are found to be responsible for the pinning of flux lines in low and high magnetic fields respectively. Our studies reveal that the choice of the composition and the heat treatment which leads to eutectic reaction are important in improving the critical current density in this alloy system. 
\end{abstract}

\maketitle

\section{Introduction}
The C15 Laves phase ZrV$_2$ and HfV$_2$ based systems have been considered to be promising materials for  high temperature applications, nuclear reactors, hydrogen storage devices\cite{wil54, zha12, cui16} as well as for high field applications alternative to the Nb based alloys and compounds\cite{ino85, ten81, tac79}. In comparison with the Nb based materials, the ZrV$_2$ and HfV$_2$ based systems were also found to be more suitable for superconducting applications in neutron radiation environment.\cite{bro77, nas84} The critical current density ($J_C$) of V/Hf$_{0.4}$Zr$_{0.6}$ composite tape has been reported to be larger than that of Nb$_3$Sn in fields higher than 14~T and at temperatures below 4.2~K.\cite{tac79} However, the C15 Laves phase materials are soft\cite{ino81} and brittle at room temperature and below,\cite{ten81} which is the limiting factor in utilizing these materials in the superconducting magnet applications. The development of new processing technologies such as powder in tube (PIT) technique, rapid heating and quenching (RHQ) and rod restack process etc., for fabricating superconducting wires continues to attract research interest in these materials.\cite{his04, seg16} There are a few reports on the routes to improve the mechanical properties of these materials such as RHQ and/or the addition of Ti and Nb.\cite{ten81, ino79} 

The PIT technique has been extensively explored to make multifilamentary wires of brittle superconducting materials such as high $T_C$ ceramics,\cite{bea97, ima18} MgB$_2$,\cite{aky07} Fe based superconductors,\cite{ma09, hua18} and V$_3$Ga,\cite{mot16} as well as Nb$_3$Sn\cite{mot16, lin00}. Similarly, brittle ZrV$_2$ based materials can also be promising candidates for making multifilamentary wires using the PIT technique. In order to progress in this direction, it is important to understand the reaction kinetics of the V$_{1-x}$Zr$_x$ solid solutions and their superconducting properties. The solubility of zirconium in vanadium at room temperature is only about 2\%.\cite{ser05} The equilibrium phase diagram of the V$_{1-x}$Zr$_x$ alloys studied by J. T. Williams\cite{wil54} shows a peritectic isotherm at 1300~$^0$C for the composition range 0.02$\leq x \leq$ 0.33 and an eutectic isotherm at 1230~$^0$C for the composition range 0.33$\leq x \leq$1. In the peritectic reaction, the solid $\beta$-V phase formed out of the melt reacts with the liquid phase and forms another solid $\gamma'$-ZrV$_2$ (C15) phase\cite{ser05}. In the eutectic reaction, the liquid being cooled transforms into two solid $\beta$-Zr and $\gamma$-ZrV$_2$ phases. Upon further cooling, the $\beta$-Zr phase transforms below 777~$^0$C into the $\alpha$-Zr phase. Therefore, by changing the composition and heat treatment schedule, it seems possible to optimize the mechanical properties and the capacity to carry large dissipationless current of wires formed by the PIT technique. In this direction, we have studied the structural, electrical and magnetic properties of arc-melted, as-cast V$_{1-x}$Zr$_x$ ($x$ = 0 - 0.40) alloys.  

Apart from the $\beta$-V and $\gamma$-ZrV$_2$ phases, the arc-melted samples in the composition range 0.38 $\leq x \leq$ 0.68 contain other additional phases such as the $\alpha$-Zr and $\beta$-Zr.\cite{cui16} The relative amount of different phases depends on the thermal treatments.\cite{cui16, fin78} The superconducting transition temperature ($T_C$) of the $\beta$-V, ZrV$_2$, and $\alpha$-Zr phases are 5.4~K, 8.8~K and 0.6~K respectively.\cite{von82} The $\beta$-Zr phase exists at high pressures and has a T$_C$ of 4-11~K depending on the applied pressure.\cite{aka90} Because of the different metallurgical phases, the V$_{1-x}$Zr$_x$ alloys with $x >$ 0.2 exhibit the signatures of multiple transitions.\cite{fin78} Superconductivity in these alloys is known to exist in high applied magnetic fields at low temperatures.\cite{yes68} However, there is no report on the critical current density in the vanadium rich V$_{1-x}$Zr$_x$ alloys and its relation to the different phases that exist in these alloys. 

In this article, we show that all the as-cast V$_{1-x}$Zr$_x$ alloys undergo successive peritectic and eutectic reactions during rapid cooling after arc melting, resulting in the formation of five structural phases. The results indicate that the V-rich V$_{1-x}$Zr$_x$ alloys with more than 10 at. \% Zr content can carry a large dissipationless current. Our results also indicate that the grain boundaries and point defects are responsible for the pinning of flux lines in low and high magnetic fields respectively.

\section{Experimental}

The V$_{1-x}$Zr$_x$ alloys ($x$ = 0-0.4) were prepared by melting 99.8+ \% purity constituent elements in an arc-furnace under 99.999 \% Ar atmosphere. The samples were flipped and re-melted six times to improve the homogeneity. The structural characterization of the samples was performed with the help of x-ray diffraction (XRD) experiments, using a Bruker D8 Advance diffractometer equipped with a Cu target (wavelength, 1.54 \AA) and a LYNXEYE detector. The measurements were performed in the 2$\theta$ range of 20-90$^0$ with a step size of 0.02$^0$ and a scanning speed of 1 sec/step. For metallography, small pieces of samples attached in moulds were ground sequentially in steps using 240, 400, 600 and 1200 grit SiC paper to obtain a flat surface. After obtaining a flat surface, the samples were polished using 6~$\mu$ m, 3~$\mu$m, 1~$\mu$m, and 0.25~$\mu$m diamond paste respectively. Then the polished surfaces of the samples were etched chemically using an aqueous solution of HF and HNO$_3$ (50:1:1) to reveal the microstructures. Scanning electron microscopy (SEM) images were taken using a NoVa NanoSEM 450 instrument (FEI Company, USA) operated at 18~kV. Compositional analysis was done using a X-Flash 6130 energy dispersive spectroscopy (EDS) attachment (Bruker, Germany) and the Esprit software. The temperature dependence of resistivity were measured in a 9~T Physical Property Measurement System (PPMS, Quantum Design, USA). The temperature and magnetic field dependence of magnetization were measured in a Superconducting Quantum Interference Device based Vibrating Sample Magnetometer (MPMS-3 SQUID VSM, Quantum Design, USA) and a 16~T VSM (Quantum Design, USA).

\section{Results and Discussion}

\begin{figure}
\includegraphics[width = 100mm]{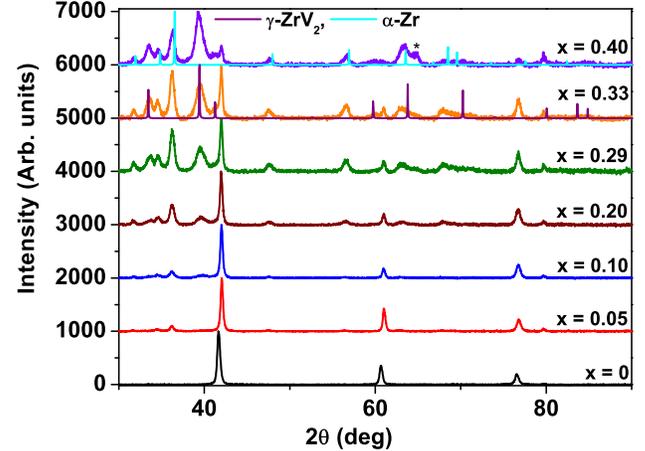}
\caption{\label{fig:epsart} (color online) X-ray diffraction patterns of the V$_{1-x}$Zr$_x$ alloys at room temperature. All the alloys contain $\beta$-V, $\alpha$-Zr and $\gamma$-ZrV$_2$ phases. The $*$ mark on the pattern for $x$ = 0.40 corresponds to the reflection from the $\beta$-Zr.}
\end{figure}

The x-ray diffraction (XRD) patterns of arc melted V$_{1-x}$Zr$_x$ alloys at room temperature are shown in Fig.1. For comparison, we have provided the XRD patterns of ZrV$_2$ and $\alpha$-Zr phases generated using the Powder Cell software \cite{kra96}. All the present alloys are formed in a multi-phase structure involving $\beta$-V, $\alpha$-Zr and $\gamma$-ZrV$_2$ phases. For $x$ = 0.40, an additional reflection (marked as $*$) from the $\beta$-Zr phase is also seen. The presence of $\alpha$-Zr phase in all the present V-rich alloys suggest that these alloys undergo both peritectic and eutectic reactions during cooling of the melt. However, the eutectic reaction takes place only when $x >$ 0.33.  Therefore, the observation of eutectic reaction in these V-rich alloys is surprising. Observation of $\beta$-Zr phase in alloys around $x$ = 0.40 indicates that the alloys at these compositions undergo eutectoid transformation below 777~$^0$C \cite{wil54}. To get more insight into the microstructure of these alloys, we have performed metallography experiments on these alloys.  

\begin{figure}
\includegraphics[width = 80mm]{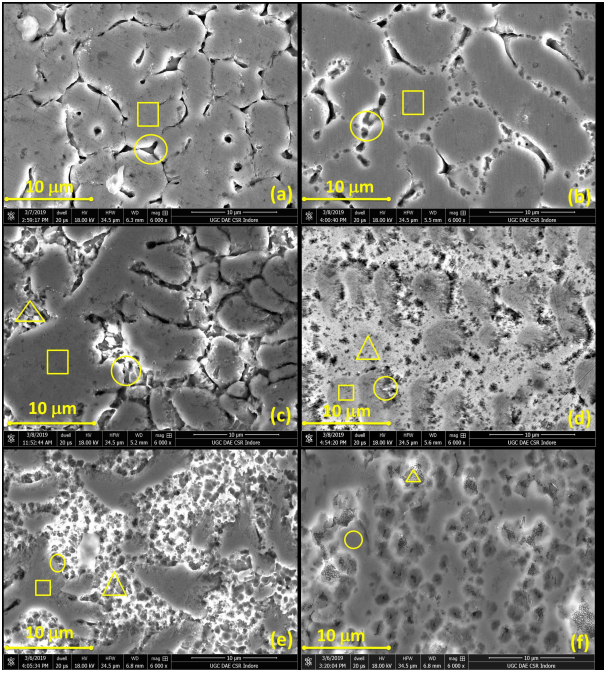}
\caption{\label{fig:epsart} (color online) Scanning electron microscopy images of polished and etched as-cast V$_{1-x}$Zr$_x$ alloys for x = (a) 0.05 (b) 0.10 (c) 0.20 (d) 0.29, (e) 0.33 and (f) 0.40. The regions similar to that marked by square has $\beta$-V structure. The regions similar to that marked by circle, C15 ZrV$_2$ ($\gamma'$) phase is formed due to peritectic reaction at 1300~$^0$C. The regions similar to that marked by triangle contains lamellar structure due to eutectic reaction at 1230~$^0$C resulting in the mixture of $\alpha$-Zr and $\gamma$-ZrV$_2$ phases.}
\end{figure}

Figure 2 shows the metallographic images of as-cast V$_{1-x}$Zr$_x$ alloys for x = (a) 0.05 (b) 0.10 (c) 0.20 (d) 0.29, (e) 0.33 and (f) 0.40. There are three different kinds of microstructures observed in these alloys. The area inside the square mark has less than 3~at.\% of zirconium. We identify this phase as $\beta$-V which solidifies first. The grain size of this phase is largest for $x$ = 0.05 and decreases with increasing $x$. Since the solubility of zirconium in $\beta$-V phase is not more than 10~at.\% even at 1300~ $^0$C, the excess zirconium content is expelled out while cooling, resulting in the zirconium enriched liquid around the solid $\beta$-V phase. This type of microstructure is observed up to $x$ = 0.33 (Fig. 2(a, b, c, d, e)). The boundaries of the $\beta$-V phase which is marked by a circle shows amount of zirconium ranging up to 50~at.\%. Here the peritectic reaction \cite{wil54, bai74} between the solid $\beta$-V phase and the zirconium enriched liquid at 1300~ $^0$C gives rise to the precipitation of $\gamma'$-ZrV$_2$ phase, resulting in further enrichment of zirconium inside the liquid. This happens due to the very short time available to the sample for this reaction to takes place during rapid cooling after the arc of the melting furnace is turned off. Now, the zirconium content in the remaining liquid crosses 33~at.\% which drives the liquid to undergo an eutectic-type reaction when the sample is cooled below 1230~ $^0$C. This reaction results in the formation of $\beta$-Zr and $\gamma$-ZrV$_2$ phases. The  $\gamma$-ZrV$_2$ and $\gamma'$-ZrV$_2$ have slightly different compositions. The $\beta$-Zr phase transforms into $\alpha$-Zr below 777~$^0$C. A small portion of the region that undergone eutectic reaction which is marked by a triangle has a rich Zr content up to 75~at.\% and shows a lamellar microstructure. While the overall microstructure of the $x$ = 0.05, 0.10 and 0.20 alloys appears to be peritectic type, hypo-eutectic type  \cite{bai74} features became more prominent at higher Zr concentration. The $\gamma$-ZrV$_2$ is preferentially etched out as it is more reactive with HF \cite{wil54}. The microstructure of the $x$ = 0.40 alloy is slightly different from those of the other alloys. The metallography images (see Fig. 2(f)) do not show any indication of the $\beta$-V phase, even though XRD results do indicate the presence in small amount. The area inside the circle has 25-45 at.\% of zirconium which is in the $\gamma'$-ZrV$_2$ phase and the area inside the triangle has about 55 at.\% of zirconium which is a mixture of $\gamma$-ZrV$_2$ and $\alpha$-Zr phases. The overall micro structure of this alloy is also hypo-eutectic type \cite{bai74}. Therefore, the non-equilibrium phase diagram of the V$_{1-x}$Zr$_x$ alloys presented here is considerably different from the equilibrium phase diagram. In literature, similar microstructures were reported for the alloys with $x \geq$ 0.43 only \cite{cui16}. 

The temperature dependence of resistivity of the V$_{1-x}$Zr$_x$ alloys in the range 2-300~K is shown in Fig.3. Addition of zirconium in vanadium increases the residual resistivity due to the formation of different phases. The inset to Fig. 3 shows an expanded view of the temperature dependence of resistivity of the V$_{1-x}$Zr$_x$ alloys near the $T_C$. Vanadium becomes superconducting below 5.2~K. The $T_C$ of the $x$ = 0.05 alloy is about 5.6~K in which the $\beta$-V phase has a composition V$_{0.98}$Zr$_{0.02}$. This alloy also shows a moderate drop in resistivity below 8.5~K due to the presence of small amount of ZrV$_2$ phase. The present alloys with $x >$ 0.05 show zero resistance just below 8.5~K. Therefore, the percolation threshold of the ZrV$_2$ phase in the V$_{1-x}$Zr$_x$ alloys lies in between $x$ = 0.05 and $x$ = 0.10. The $T_C$ of all the alloys are estimated as that temperature where the temperature derivative of resistivity shows a peak. These values are presented in Table 1.      

\begin{figure}
\includegraphics[width = 80mm]{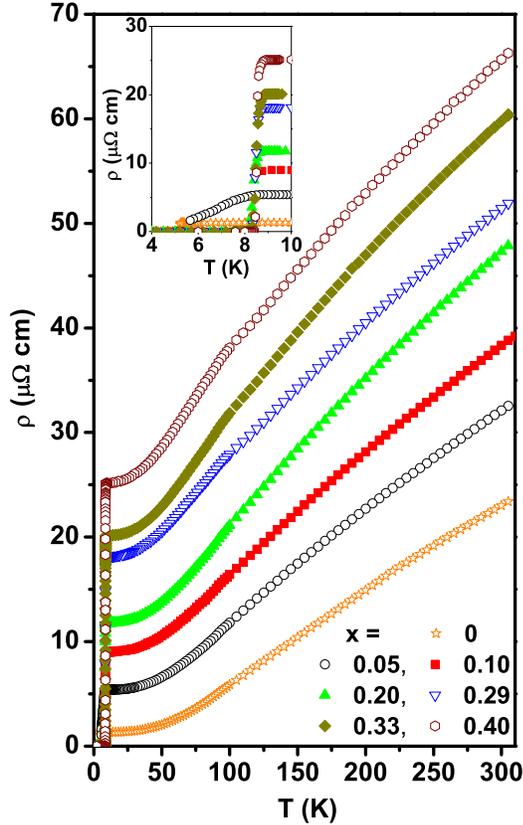}
\caption{\label{fig:epsart}(color online) Temperature dependence of electrical resistivity of the V$_{1-x}$Zr$_x$ alloys. Addition of zirconium in vanadium increases the residual resistivity due to the formation of secondary phases. (Inset) Expanded view of the temperature dependence of resistivity of the V$_{1-x}$Zr$_x$ alloys near the superconducting transition. For all the alloys with $x >$ 0.05, the transition from normal state to superconducting state takes place below $\sim$8.5~K indicating that these alloys have crossed the percolation threshold for the ZrV$_2$ phase in the V$_{1-x}$Zr$_x$ alloys.}
\end{figure}

\begin{figure}
\includegraphics[width = 80mm]{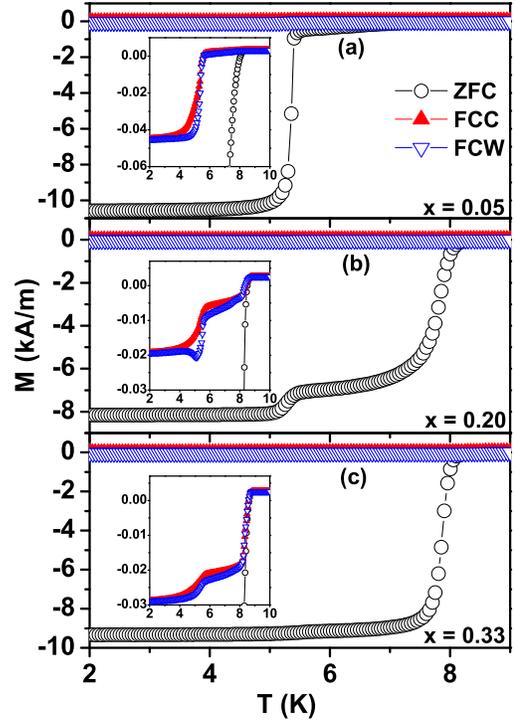}
\caption{\label{fig:epsart}(color online) Temperature dependence of magnetization of the V$_{1-x}$Zr$_x$ ($x$ = 0.05, 0.20 and 0.33) alloys in the range 2-9~K measured in 10~mT magnetic field. All the samples above $x >$ 0.05 exhibit a two-step transition when the temperature is decreased below 9~K. This is due to the superconductivity in ZrV$_2$ ($T_C$ = 8.5~K) and $\beta$-V ($T_C$ = 5.2~K) phases.The insets show the expanded view of the data measured in FCC and FCW protocols.}
\end{figure}

Figure 4 shows the temperature dependence of magnetization ($M$($T$)) measured in 10~mT magnetic field while warming-up the sample after initially cooling it down to 2~K in the absence of external magnetic field (ZFC), during cooling the sample down to 2~K in the presence of 10~mT from a temperature well above the $T_C$ (FCC), and during warming-up in the presence of 10~mT field after the FCC run (FCW). The $M$($T$) curves exhibit a two-step transitions in all the samples with $x >$ 0.05, when the temperature is decreased below 9~K. This indicates the expulsion of the magnetic flux due to the superconductivity in the ZrV$_2$ ($T_C$ = 8.5~K) and $\beta$-V ($T_C$ = 5.2~K) phases. The ratio of the superconducting volume fractions of the ZrV$_2$ and $\beta$-V phases can be estimated from the ZFC magnetization response from the shielding current induced when a small magnetic field less than the lower critical field is applied in the superconducting state. We have estimated the superconducting volume fraction ($S_f(ZrV_2)$) of the ZrV$_2$ phase with respect to $\beta$-V phase as M(5.7~K)/M(2~K), and this is given in table 1. The $S_f(ZrV_2)$ increases with increasing $x$ which indicates that for the present alloys with $x >$ 0.05, the major superconducting phase is ZrV$_2$. The $S_f(ZrV_2)$ is not 100\% for any of the present alloys which indicates that the $\beta$-V phase  exists all the way up to $x$ = 0.40.   The insets to Fig. 4 show that the magnitude of $M$ during the FCC measurements is quite small as compared to that during the ZFC measurements. This indicates that the flux line pinning is strong in these alloys. 

\begin{table} 
\caption{\label{tab:table1} Residual resistivity, $T_C$ and the superconducting volume fraction $S_f$ for various V$_{1-x}$Zr$_x$ alloys}
\begin{tabular}{cccccccc}
\hline
x &residual resistivity&T$_C$&S$_f$(ZrV$_2$)\\
&($\mu \Omega$ cm)&(K)&(\%)\\
\hline
0& 1.28&5.2 & 0  \\
0.05&5.39& 5.6& 4  \\
0.10& 8.99&8.4& 89  \\
0.20&11.73& 8.3 & 87 \\
0.29& 18.11&8.5 & 96  \\
0.33&20.1& 8.5& 98  \\
0.40&25.1& 8.5 & 98.8 \\
\hline
\end{tabular}

\end{table}

\begin{figure}
\includegraphics[width = 80mm]{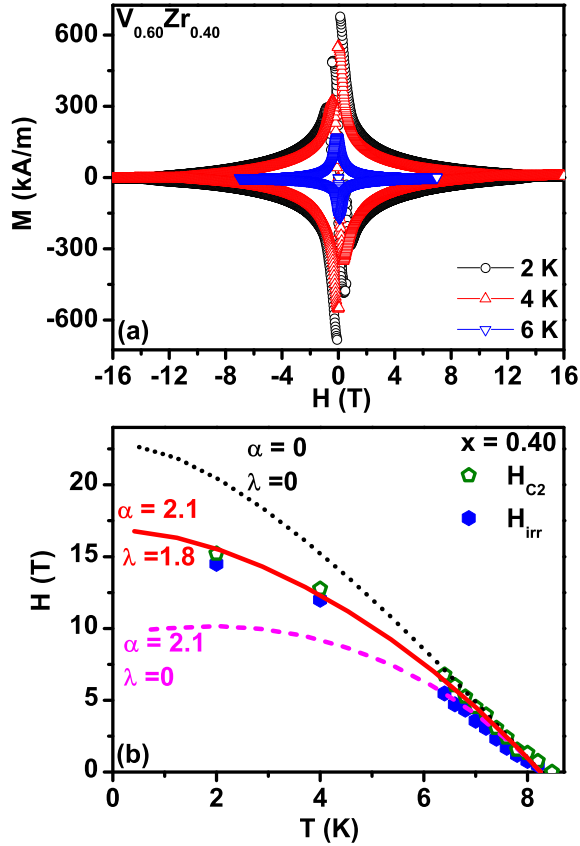}
\caption{\label{fig:epsart}(color online)  (a) Magnetic field dependence of magnetization of the V$_{0.60}$Zr$_{0.40}$ alloy at 2~K, 4~K and 6~K. (b) The temperature dependence of $H_{C2}$ and H$_{irr}$ of the V$_{0.60}$Zr$_{0.40}$ alloy. The symbols represent the experimental data and the lines are the fits obtained using the WHH formalism. The $H_{C2}$(T = 0) is about 17.5~T.}
\end{figure}

Figure 5(a) shows the magnetic field dependence (up to 16~T) of magnetization of the V$_{0.60}$Zr$_{0.40}$ alloy. The hysteresis in M(H) between the increasing and decreasing fields closes at the irreversibility field H$_{irr}$ = 15~T at 2~K. Figure 5(b) shows the temperature dependence of upper critical field (H$_{C2}$) and H$_{irr}$ of the V$_{0.60}$Zr$_{0.40}$ alloy. The magnetic field at which a distinct deviation takes place from the magnetic field dependence of the normal state magnetization is taken as the H$_{C2}$. This procedure has been effectively used to estimate the H$_{C2}$ in superconductors such as borocarbides \cite{roy94, roy96} and skutterudies \cite{sha12} where an enhanced paramagnetism is observed in the normal state. The error in the estimation of H$_{C2}$ is found to be less than 2~\%. We have observed that the H$_{C2}$($T$) of all the alloys with $x >$ 0.05 are of similar magnitude. The symbols represent the experimental data points. The value of $H_{C2}$(4.2~K) is similar to the reported value of experimental $H_{C2}$(4.2~K) obtained from the resistivity measurements \cite{yes68}. The lines shows the fits obtained using the Werthamer, Helfand and Hohenberg (WHH) model \cite{wer66}. The dotted line (black) which represents the estimated H$_{C2}$($T$) when the paramagnetic effects and spin-orbit coupling are not considered, is higher than the experimental data points at low temperatures. This suggest that the paramagnetic effects are important in these alloys. The dashed line (pink) is estimated by considering the paramagnetic effects alone ($\alpha$ = 2.1). This curve lies below the experimental data at low temperatures indicating the presence of spin orbit coupling ($\lambda \neq$ 0). The solid line (red) which is best fitted curve is obtained for $\alpha$ = 2.1 and $\lambda$ = 1.8.

\begin{figure}
\includegraphics[width = 80mm]{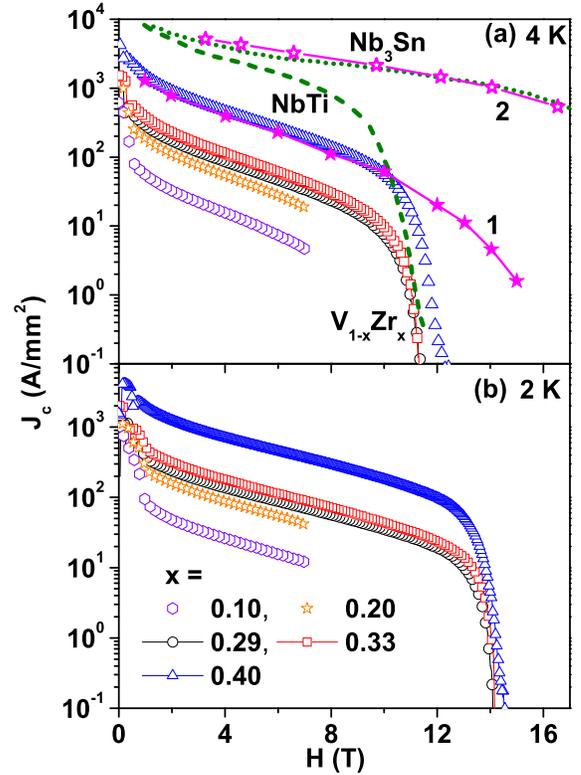}
\caption{\label{fig:epsart}(color online) Magnetic field dependence of critical current density of the V$_{1-x}$Zr$_x$ alloys for $x$ = 0.10, 0.20, 0.29, 0.33 and 0.40 at (a) 2~K and (b) 4~K. For comparison, we have plotted the $J_C$(H) of modern NbTi wires (dashed line), Nb$_3$Sn wires (dotted line),  (Hf,Zr)/V/Ta Multi wire (curve 1: solid star) and V-1at.\% Hf/Zr-40 at.\% Hf annealed 950~$^0$C for 200~ hr (curve 2: open star) at 4.2~K.}
\end{figure}

\begin{figure}
\includegraphics[width = 80mm]{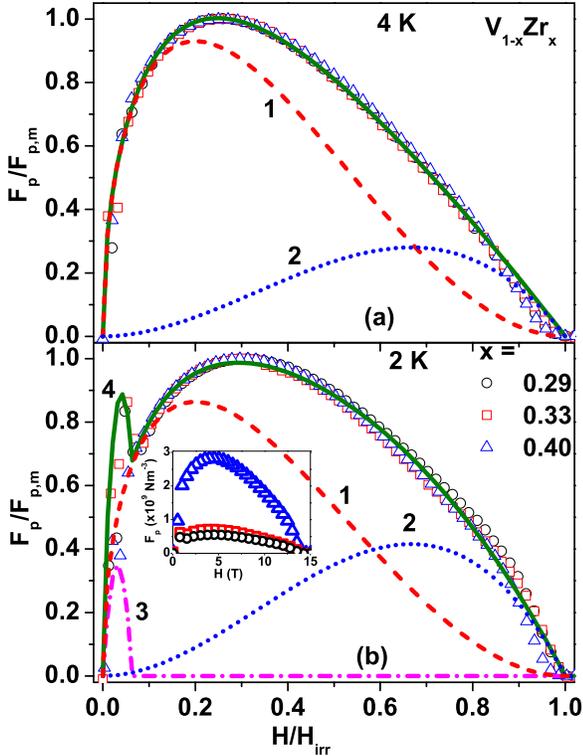}
\caption{\label{fig:epsart}(color online) Normalized pinning force density of the V$_{1-x}$Zr$_x$ alloys as a function of $h$ at (a) 2~K and (b) 4~K. It is observed that grain boundaries (curve 2) and point defects (curve 3) pin the vortex lines at low and high magnetic fields respectively. (Inset) Magnetic field dependence of pinning  force density  at 2~K indicates that the pinning strength of the $x$ =0.4 alloy is more than 5 times higher than other alloys.}
\end{figure}

The magnetic field dependence of critical current density ($J_C$($H$)) of the V$_{1-x}$Zr$_x$ alloys estimated using $J_C$ = 2$\Delta M$/$\left[a(1-a/3b)\right]$ is shown in Fig.6. Here, the $\Delta M$ is estimated from the $M$($H$) using the critical state model as $\Delta M$ = $M$($H$ decreasing) - $M$($H$ increasing), $a$ is the width and $b$ is the breadth of sample and $a < b$\cite{shy15}. The cross sectional area $ab$ is perpendicular to the direction of applied magnetic field. Initially, we have measured the magnetization of all the samples up to 7~T and estimated the $J_C$($H$). We found that the $J_C$($H$) increases with increasing $x$, and for alloys with $x >$ 0.10, the $J_C$($H$) lies in the range 10$^2$-10$^3$~A/mm$^2$. We have also found that the $J_C$($H$) is quite high even at 7~T. Therefore, we have estimated the $J_C$($H$) of $x$ = 0.29, 0.33 and 0.40 alloys up to 16~T at 2~K and 4~K. The $J_C$(0) of these three alloys is more than 10$^3$~A/mm$^2$ and decreases with increasing field. Amongst the present alloys, the V$_{0.60}$Zr$_{0.40}$ alloy is observed to have highest $J_C$ in all magnetic fields at any temperature. For comparison we have plotted the $J_C$($H$) of modern NbTi wires (dashed line) \cite{end73, che91, bou06, miy06, lin15}, Nb$_3$Sn wires (dotted line)\cite{ino85, lin15, flu08},  (Hf,Zr)/V/Ta Multi wire (curve 1: solid  star) \cite{his06} and V-1at.\% Hf/Zr-40 at.\% Hf annealed 950~$^0$C for 200~ hr (curve 2: open star) \cite{ino79} at 4.2~K taken from the Ref.\cite{lin15, his06, ino79}. One can see that the $J_C$($H$) of the V$_{0.60}$Zr$_{0.40}$ alloy is about 5 times smaller than that of modern NbTi wire below 11~T and 4~K, however, is similar to a multifilamentary wire of (Hf,Zr)/V/Ta configuration prepared using RHQ technique  (closed star) \cite{his06}. Moreover, $J_C$ is significant in fields above 11~T for this alloy. The specially prepared C15 Hf-Zr-V alloys have a $J_C$ higher than 10$^3$~A/mm$^2$. The Hf-Zr-V metallic glasses have a $J_C$ of 5~$\times$~10$^3$~A/mm$^2$ at 4.2~K and 6.5~T \cite{ten81}, whereas V/(Hf$_{0.4}$Zr$_{0.6}$) composite tapes have a $J_C$ of 1~$\times$~10$^3$~A/mm$^2$ at 4.2~K and 15~T (open star in Fig.6) \cite{ino76}. However, arc-melted Hf-Zr-V samples have a $J_C$ of 2~$\times$~10$^2$~A/mm$^2$ or less at 4.2~K and 12~T \cite{ino71}. Thus, the results indicate that the present alloys are promising candidates as a probable alternate to the Nb based alloys and compounds for high magnetic applications and calls for the employment of modern processing techniques for the optimization of mechanical properties and $J_C$. 

In order to understand the role of microstructure on the high $J_C$($H$) of these alloys, we have estimated the field dependence of the pinning force density ($F_p$(H)) of the above alloys at 4~K and 2~K. The $F_p$ is estimated as $F_p = J_C \times H$. Figure 7 shows the normalized pinning force density F$_p$/F$_{p,m}$ (F$_{p,m}$ is the maximum F$_{p}$ at a given temperature) as a function of reduced field $h$ = H/H$_{irr}$ ((a) at 4~K \& (b) at 2~K). The values of F$_p$/F$_{p,m}$($h$) is almost same for all the alloys at a given temperature. This indicates that all the alloys contain similar kinds of pinning centres. The F$_p$/F$_{p,m}$ has a maximum at about $h_m \sim$~0.25 for both 4~K and 2~K.  However, F$_p$/F$_{p,m}$ decreases slowly as $h$ increases. It is also observed that there is a peak like structure at very low $h$ in few alloys at 2~K. The functional form of F$_p$/F$_{p,m}$($h$) corresponding to different kinds of pinning centres established by Dew-Hughes is presented in table 2.  \cite{dew74, mat13}. The comparison of experimental $h_m$ to that corresponding to different pinning mechanisms (table 2) indicate that the most of the flux lines are pinned by the core interaction with the normal surface pinning centres ($h_m$ = 0.2: curve 1 of Fig. 7) which has the functional form $h^{0.5}(1-h)^{2}$. Generally, these pinning centres are the grain boundaries. The experimental F$_p$/F$_{p,m}$($h$) values at $h > h_m$ are quite large than the curve 1 and also shows a change of slope at about $h \sim$ 0.8. This indicates that the pinning mechanism at high magnetic fields is different from that at low magnetic fields. The table 2 suggest that the $\Delta \kappa$ pinning due to point defects results in the maximum F$_p$/F$_{p,m}$($h$) at $h$ = 0.67 (curve 2 of Fig.7). We found that both these mechanisms of pinning need to be considered to account for the pinning properties over complete range of $H$. The low field peak observed at 2~K could not be explained by functional forms given in table 2 unless we take different $H_{C2}$/$H_{irr}$ ($\sim$ 1.2~T) in estimating $h$. The $\beta$-V phase has a value of $H_{C2}$ of about 1.2~T at 2~K. Therefore, we identify that the origin of this peak is due to the pinning of flux lines in the $\beta$-V phase which becomes superconducting below 5.4~K. Therefore, the signature for flux line pinning is nearly absent in  F$_p$/F$_{p,m}$($h$) at 4~K. We found that the flux pinning in the $\beta$-V phase is by $\Delta \kappa$ pinning due to volume pins which has the fornctional form $h(1-h)$. 

\begin{table} 
\caption{\label{tab:table1} The functional form of F$_p$/F$_{p,m}$($h$) corresponding to different kinds of pinning centres}
\begin{tabular}{cccccccc}
\hline
Geometry &Type of&Functional&Position of \\
of the pin&pinning centre&form& the maximum \\
&&&($h_m$)\\
\hline
Magnetic interaction\\
\hline
Volume&Normal&$h^{0.5}(1-h)$ & 0.33  \\
Volume&$\Delta \kappa$&$h^{0.5}(1-2h)$& 0.17,1  \\
\hline
Core interaction\\
\hline
Volume&Normal&$(1-h)^{2}$& -  \\
Volume&$\Delta \kappa$&$h(1-h)$& 0.5 \\
Surface&Normal&$h^{0.5}(1-h)^{2}$& 0.2  \\
Surface&$\Delta \kappa$&$h^{1.5}(1-h)$&0.6  \\
Point&Normal&$h(1-h)^{2}$& 0.33 \\
Point&$\Delta \kappa$&$h^{2}(1-h)$& 0.67 \\
\hline
\end{tabular}

\end{table}

The inset to the Fig. 7(b) shows the $F_p$(H) of the V$_{1-x}$Zr$_x$ alloys with $x$ = 0.29, 0.33 and 0.40 at 2~K. The $F_p$ increases with increasing $x$. The highest $F_p$ at 2~K is observed for the V$_{0.60}$Zr$_{0.40}$ alloy which is about 3~$\times$~10$^9$~Nm$^{-3}$ at about 4.5~T. The increase of F$_p$ with increasing $x$ indicates the increase in the amount of pinning centres. We have observed that the main envelop can be fitted by taking F$_p$/F$_{p,m}$ =A $h^{0.5}(1-h)^{2}$ + B $h^{2}(1-h)$ where A is the weight factor for strength of pinning of flux lines by the grain boundaries and B is that for the point defects. The ratio A/B $>$ 1 indicates that the grain boundaries are the predominant pinning centres. The metallography images suggest that the large amount of grain boundaries are generated due to the formation of $\gamma$-ZrV$_2$ and $\beta$-Zr (lamellar structure) from the eutectic type reaction of the zirconium enriched liquid below 1230~$^0$C. We could not find the $\beta$-Zr phase in the metallography images. This suggests that the size of the $\beta$-Zr phase is quite small and is present inside the $\alpha$-Zr phase. Thus, $\beta$-Zr phase may act as point defects. The area under the curve 2 (pinning due to point defects) increases and that of the curve 1 (pinning due to grain boundaries) decreases with the decreasing temperature. Therefore with the decreasing temperature, the strength of the flux line pinning by the grain boundaries decreases in comparison to the strength of the flux line pinning by the point defects. Therefore, the choice of composition and the thermal treatments that leads to eutectic type reaction is important in improving the $J_C$($H$) in these alloys. 

\section{Conclusions}
In conclusion, we have studied the structural, electrical and magnetic properties of the arc melted as-cast V$_{1-x}$Zr$_x$ alloy superconductors. From the comparison of our metallography results with the existing equilibrium phase diagram, we conclude that when the molten V$_{1-x}$Zr$_x$ alloys are cooled rapidly, \\
(i) the vanadium rich region solidifies below $\sim$ 1600~$^0$C. This leads to the expulsion of excess Zr from the solid $\beta$-V phase and a Zr rich liquid forms along the grain boundaries. \\
(ii) When the temperature is reduced below 1300~$^0$C, the surface of $\beta$-V reacts with the Zr rich liquid (L1) to form a layer of $\gamma'$ phase (peritectic reaction). The formation of $\gamma’$ phase again enriches the Zr content in the remaining liquid (L2). \\
(iii) Upon further reduction of temperature below 1230~$^0$C, the L2 liquid phase separate into $\gamma$ and $\beta$-Zr phases (eutectic type reaction).\\
(iv) Below 777~$^0$C, the $\beta$-Zr phase transforms into the $\alpha$-Zr phase. However, depending on the cooling rate, there can be precipitation of untransformed $\beta$-Zr phase as well. \\

 Our studies also show that the percolation threshold for bulk superconductivity due to the ZrV$_2$ phase in the V$_{1-x}$Zr$_x$ alloys is less than 10 at.\% of zirconium in vanadium. The $J_C$ of the V$_{1-x}$Zr$_x$ alloys with $x >$0.29 is in the range of modern Nb-Ti wires. The analysis of the normalized flux pinning force density suggest that by driving the alloy to undergo an eutectic type reaction, we can increase the capacity of the material to carry large dissipationless current. Thus, the choice of the composition and the heat treatment that leads to the eutectic type reaction is important in improving the $J_C$($H$) in these alloys.

\vskip -1 cm
\nocite{*}
\bibliography{aipsamp}

\end{document}